\documentclass[{article}
 reprint,
 amsmath,amssymb,
 aps,
]{revtex4-2}
\usepackage{setspace}
\usepackage{graphicx}
\usepackage{bm}
\usepackage{amsthm}

\def\be{\begin{equation}}
\def\ee{\end{equation}}
\def\a{\alpha}

\def\be{\begin{equation}}
	\def\ee{\end{equation}}
\def\ba{\begin{eqnarray}}
	\def\ea{\end{eqnarray}}
\def\la{\langle}
\def\ra{\rangle}
\def\a{\alpha}

\def\h{\hskip 1cm}

\def\lo{\longrightarrow}
\newtheorem{theorem}{Theorem}
\begin{document}

\preprint{APS/123-QED}

\title{Capacities of the covariant Pauli channel }
\author{Abbas Poshtvan$^{1}$}
\author{Vahid Karimipour$^{1}$}%
\affiliation{%
 $^1$Deptartment of Physics, Sharif University of Technology, Tehran14588 , Iran
}%




\date{\today}

\begin{abstract}
 We study four well-known capacities of a two-parameter family of qubit Pauli channels. These are the channels which are covariant under the $SO(2)$ group and contain the depolarizing channel as a special case. We find exact expressions for the classical capacity and entanglement assisted capacities, and analytically determine the regions where the quantum capacity of the channel vanishes. We then use a flag extension to find upper bound for the quantum capacity and private capacity of these channels in the entire region of parameter space and also obtain the lower bound for the quantum capacity by calculating the single shot quantum capacity numerically.  In conjunction with previous results on depolarizing channels, our result is one step forward for determining the capacities of the full Pauli channel.  
\end{abstract}

\maketitle


\section {Introduction}
For a wide range of tasks related to storing, processing and communicating information, protection of  the careers of information from environmental noise is a central issue. Many efforts have been made to realize and implement the techniques of this protection and also to understand the fundamental limits of our ability to prevent the information from being ruined by the noises. In the quantum domain, where classical or quantum information are encoded into quantum states, the effect of environmental noise is  modeled by a Completely Positive Trace-Preserving (CPTP) map or quantum channel for short. The capacity of a channel then determines the ultimate rate at which a sender can reliably transmit information to a receiver over that channel, where it is taken into account that the sender and the receiver can respectively use arbitrary encoding and decoding procedures. \\

In contrast to classical information theory where there is a unique definition of classical capacity \cite{shannon1948mathematical}, in the quantum domain, depending on the auxiliary resources available (e.g. shared entanglement and classical communication), allowed types of encoding and decoding (e.g separable or entangled encoding) and the types of information (e.g. classical or quantum) which is to be transferred  \cite{shor2002quantum}, we encounter different operational definitions of capacity. There is usually a long route from these operational definitions to closed formulas for determination of these capacities. 
Generally for a quantum channel $\Lambda$,  the final expression of these capacities take the form 
\be
C_r(\Lambda)=\lim_{n\lo \infty}\frac{1}{n}\Xi_r(\Lambda^{\otimes n})
\ee
where the quantity $\Xi_r$ depends on the type of capacity $r$ which is to be calculated and $\Xi_r(\Lambda)$ itself is obtained by an optimization over an ensemble of states.
In other words for each kind of capacity, the function $\Xi_r$ takes a specific form derived from the operational definition of that kind of capacity. 
 Even then, calculation of these closed formulas are not always tractable analytically, except in some special cases \cite{smith2008quantum,ouyang2011channel, kianvash2020bounding, fanizza2021estimating, fanizza2020quantum, smith2008additive, chessa2021quantum, d2013classical, arqand2020quantum, filippov2019quantum, oskouei2021capacities, leditzky2017quantum}. In fact the regularized capacity should be denoted by 
 $C_r^\infty$ to emphasize the necessity of the limiting procedure, however we follow the usual nomenclature and denote it simply by $C_r$.  The regularization, i.e. the limit $n\longrightarrow \infty$ is necessary since in the encoding and decoding procedures at the beginning and end of the channel, one can in principle use infinitely long sequences of entangled states and entangled measurements respectively. Needless to say, this regularization procedures is extremely difficult if not impossible. The superadditivity problem refers to the situation where entangled input states and entangled measurement can increase a specific type of  capacity of a channel. When superadditivity holds, it means that  $\Xi_r(\Lambda^{\otimes n})\ne n\Xi_r(\Lambda)$ and hence there is no way to bypass 
the limiting procedure $n\lo \infty$. Investigation of  superadditivity  and other properties like superactivation, and causal activation \cite{calefi1, calefi2}, which have no classical counterpart comprises an important chapter in the field of quantum information theory with many inter-related contributions, a chapter which is still growing .
The main question of additivity is whether or not a certain quantity $\Xi$ adds up when we take the tensor product of two quantum channels, i.e. whether or not
\begin{equation}\label{add1}
\Xi_r (\Lambda_1\otimes \Lambda_2)\stackrel{?}{=}\Xi_r(\Lambda_1)+\Xi_r(\Lambda_2).
\end{equation}
The quantity $\Xi_r$ can take different forms, depending on the context, i.e. it can be the minimum output entropy, or it can be the Holevo quantity or it can be 
a generalization of Holevo quantity, where the Shannon entropy $S(\rho):=-Tr(\rho \log \rho)$ is replaced \cite{R3} by the Renyi entropy $S_p(\rho):=\frac{Tr\rho^p}{1-p}$. 
In an extensive effort for the study of additivity conjectures, it has been found that some of these conjectures are indeed equivalent to each other \cite{R27} and some others have been disproved by finding counter-examples \cite{hastingnature, fukudasimplification,Fukuda, FukudaWolf, R28, R14, R7}.
The general result is that for a  generic quantity $\Xi$ and a generic channel $\Lambda$, the additivity property does not hold and only in certain cases, one can benefit the additivity property to calculate certain capacities of certain channels \cite{R1,R2,R6,R9,R17,R18,R19,R20}. For a review see (\cite{fukreview}). In most of the other cases, one has to suffice to compute upper and lower bounds for the capacity. We will see examples of this in the sequel. \\

Even if one restricts oneself  to calculation of single shot capacites, $C^1_r$,  one is still faced with an obstacle which  is the absence of concavity of $\Xi(\Lambda)$ for certain functions. In this case there is no simple clue for finding the global maximum and hence one should perform  an optimization over a very large set of parameters  \cite{devetak2005private} \cite{hastings2009superadditivity}.\\

\noindent It is therefore no surprise that for a qubit channel as simple and as ubiquitous as the  Pauli channel \be\label{gPalui}\Lambda(\rho)=p_0\rho+p_1 X\rho X+p_2 Y\rho Y + p_3 Z\rho Z,\ee  nontrivial results exist only for special one-parameter classes, like the depolarizing channel \cite{fanizza2020quantum} which has only one parameter. It is therefore desirable to calculate these capacities for a wider class of channels, which due to their symmetry properties allow exact calculations. One can then hope that these results, together with continuity property of the capacity, will provide insight for the capacities of the full 3-parametric Pauli channel.\\

In this paper, we analyse the classical, entangled assisted, quantum and private capacities of the covariant Pauli channel. 
This is a subclass of Pauli channels defined as 
\begin{equation}\label{Channel}
	\Lambda(\rho)=p_0\rho+p_1(X\rho X+Y\rho Y)+p_3Z\rho Z,
\end{equation}
subject to the trace-preserving condition, $p_0+2p_1+p_3=1$. In contrast to the depolarizing channel which is covariant under the full $SU(2)\sim SO(3)$ rotations, this channel is covariant only under rotations around the third axis, i.e.
\begin{equation}
	\Lambda(U_z(\theta)\rho U^\dagger_z(\theta))=U_z(\theta)\Lambda(\rho)U^\dagger_z(\theta),
\end{equation}
where $U_z(\theta)=e^{i\theta\frac{Z}{2}}$ is an arbitrary  rotation around the $z$ axis. 
Calculation of  capacities of this channel will bring us one step closer to the calculation of the general Pauli channel \cite{leung2009continuity} and in fact every qubit channel. The latter statement is the result of a theorem  \cite{horodecki1999general} according to which any qubit channel can be twirled to a Pauli channel with no-higher quantum capacity.   \\

\noindent The structure of this paper is as follows: In section (\ref{Comp}), we review the concepts of complement of a general quantum channel,  and relate the covariance properties of a channel with that of its complementary channel.  In section (\ref{Covpauli}) we restrict ourselves to the qubit Pauli channel and analyse in detail its $U(1)\sim SO(2)$ covariance properties. In particular this study reveals some symmetries in the spectrum of the output states for both the channel and its complement which will play a crucial role in determining exact expressions for the capacities.    Finally, in section (\ref{capacitiysection}) we will investigate these capacities for the covariant Pauli channel. In particular, we exactly calculate the classical and entanglement-assisted capacities and use flag extension to find upper and lower bounds for the quantum and private capacity of this channel.  The paper ends with a discussion.\\

\section{The complement of a general quantum channel}\label{Comp}
Calculation of many types of capacities reduces to optimization of quantities related to both the channel and its complement. Therefore in this subsection we review the concept of the complement of a channel and investigate how the covariance and symmetry properties of a channel are reflected in its complement. \\

\begin{figure}
     \centering
     \includegraphics[width=10.0 cm,height=5cm,angle=0]{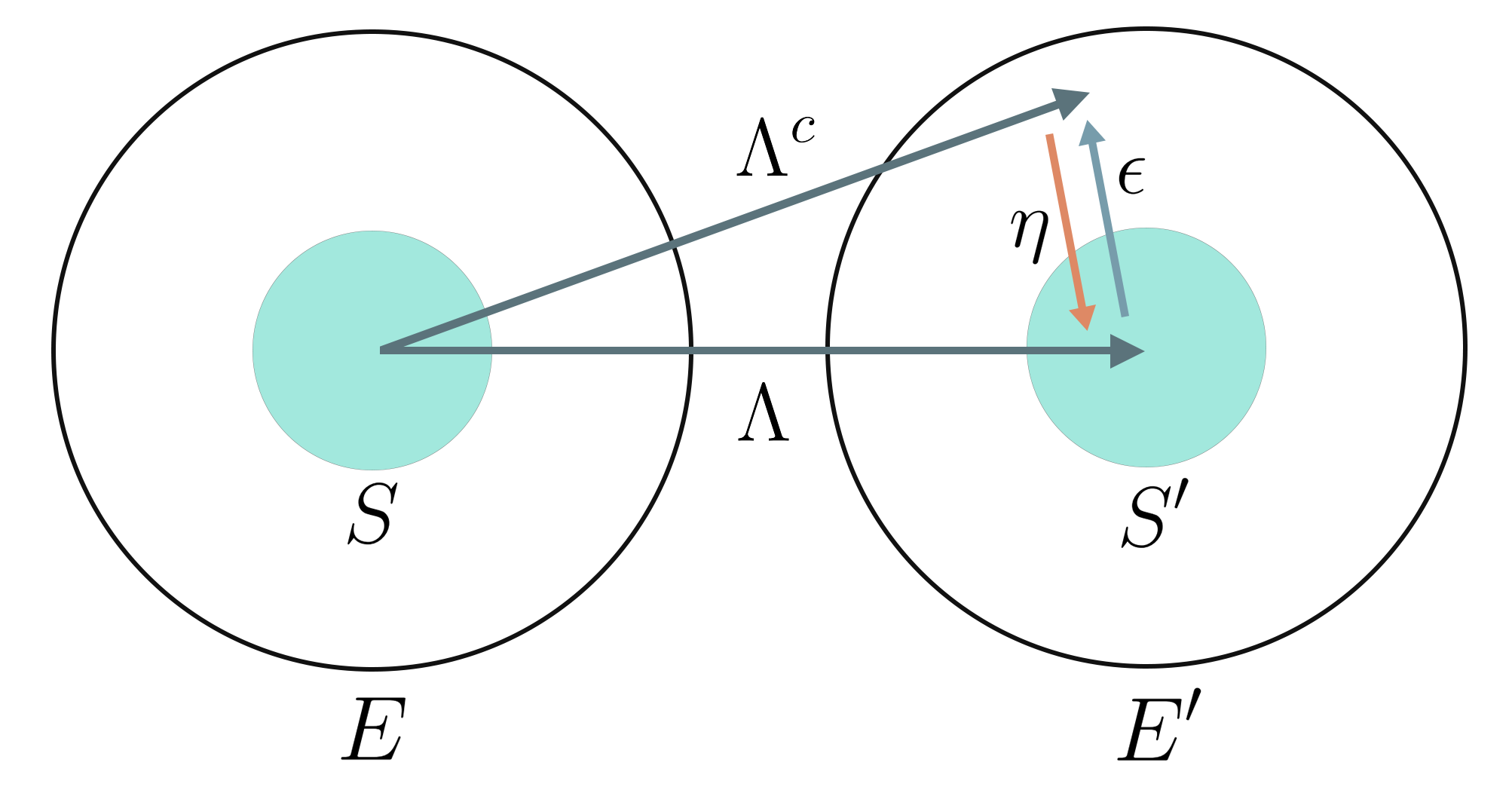}\vspace{-0.1cm}
     \caption{(Color Online) Spaces of the input and output systems ($S,S'$) and environments ($E,E'$) and the channels transforming their states ($\Lambda, \Lambda^c$) (eqs. (\ref{steinch}) and (\ref{steincomp})) .}
     \label{Spaces}
 \end{figure}

\noindent For a Hilbert space $H$, we denote by $L(H)$ the space of linear operators on $H$ and by $L^+(H)$  the set of positive operators on $H$. A quantum channel is a completely positive trace preserving map with input system $S$ and output system $S^{'}$ $\Lambda: L^+(H_S) \longrightarrow L^+(H_{S^{'}})$. If we denote the environment of the input and output systems as $E$ and $E^{'}$ as in figure (\ref{Spaces}), we can write the channel's expression in the Stinespring form:
\begin{equation} \label{steinch}
	\Lambda(\rho)= tr_{E'} (V \rho V^{\dagger})
\end{equation}
where V is an isometry from $H^+_S$ to $H^+_{S^{'}} \otimes H^+_{E^{'}} $. In this setup, the complementary channel $\Lambda^c: L^+(H_S) \longrightarrow L^+(H_{E^{'}})$ is: 
\begin{equation} \label{steincomp}
	\Lambda^c(\rho)= tr_{S'} (V \rho V^{\dagger})
\end{equation}
which is a map from the input system to the output environment (figure \ref{Spaces}). The complement of a quantum channel is not unique, but they are connected to each other by isometries \cite{fukdatta}. The Kraus operators of the channel $\Lambda$ and its complement $\Lambda^c$ are related as follows
\cite{smaczynski2016selfcomplementary}: 
\begin{equation} \label{krauscomp}
	\begin{split}
		&\Lambda(\rho)= \sum_\alpha K_{\alpha} \rho K_{\alpha} ^{\dagger} \\
		&\Lambda^c(\rho)= \sum_i R_i \rho R_i ^{\dagger} \\
		&(R_i)_{\alpha,j}= (K_{\alpha})_{i,j}
	\end{split}     
\end{equation}

\noindent {\bf Degradable and Anti-degradable Channels:}
The channel $\Lambda$ is called degradable if there exists another CPTP map $\epsilon: L^+ (H_{S^\prime}) \longrightarrow L^+ (H_{E\prime})$ such that $\epsilon \circ \Lambda=\Lambda^c $. On the other hand, $\Lambda$ is said to be anti-degradable if there exists a CPTP map $\eta: L(H_{E^\prime}) \longrightarrow L(H_{S^\prime})$  such that $\eta \circ \Lambda^c=\Lambda$. (figure \ref{Spaces}). 
A natural question is that what type of covariance is induced on the complement of a quantum channel, by the covariance of the channel itself.  We present the answer to this question in the following theorem. \\
\begin{theorem}\label{Glambdac}
Let a general quantum channel $\Lambda(\rho)=\sum_\alpha K_\alpha \rho K_\alpha^\dagger $ be $G-$ covariant in the sense that 
\be
\Lambda \big[U(g)\rho U^\dagger(g)\big]=V(g)\Lambda(\rho)V^\dagger(g)\h \forall \ g\in G,
\ee
where $U(g)$ and $V(g)$ are two representations of the group element $g$ in a group $G$. Then the complement channel is covariant in the following form
\be \label{covofcomp}
\Lambda^c\big[U^\dagger(g)\rho U(g)\big]=\Omega^\dagger (g)\Lambda^c(\rho)\Omega(g)\h \forall \ g\in G,
\ee
where $\Omega(g)$ is the representation defined in the following form
\be\label{cov1}
V^\dagger(g)K_\alpha U(g)=\sum_{\beta}\Omega_{\alpha,\beta}(g)K_{\beta}.
\ee
\end{theorem}
\begin{proof}
We start from (\ref{cov1}) which is a consequence of the covariance of the channel $\Lambda$ and use the relation (\ref{krauscomp}) between the Kraus operators of the channel and its complement:
\be
(R_{i})_{\a,j}\equiv (K_\a)_{i,j}=\sum_{\beta}\Omega_{\a,\beta}(V K_\beta U^\dagger)_{i,j}.
\ee
Expanding the right hand side and once again using (\ref{krauscomp}), we find
\ba
(R_{i})_{\a,j}&=&\sum_{\beta,m,n}\Omega_{\a,\beta}V_{im}(K_\beta)_{m,n}{U^\dagger}_{n,j}\cr
&=&\sum_{\beta,m,n}\Omega_{\a,\beta}V_{im}(R_m)_{\beta,n}{U^\dagger}_{n,j}\cr
&=&\sum_{m}(\Omega R_mU^\dagger)_{\a,j}{V}_{i,m}.
\ea
Inserting this in the Kraus representation of the complement channel, we find
\ba
\big[\Lambda^c(\rho)\big]_{\a,\beta}&=&\sum_i (R_i)_{\a,j}\rho_{jk}\overline{(R_i)_{\beta,k}}=\sum_{i,m,n}(\Omega R_m U^\dagger)_{\a,j}V_{i,m}\rho_{jk}\overline{(\Omega R_n U^\dagger)_{\beta,k}}\overline{V_{i,n}}.
\ea
Using the unitarity of $V$, the right hand side is simplified and we find 
\be
\big[\Lambda^c(\rho)\big]_{\a,\beta}
=\sum_{m,n}(\Omega R_m U^\dagger)_{\a,j}\rho_{jk}(U R_m^\dagger \Omega^\dagger)_{k,\beta},
\ee
which leads to the final result, namely
\be
\Lambda^c(\rho)=\Omega \Lambda^c(U^\dagger\rho U)\Omega^\dagger.
\ee
This proves the theorem.
\end{proof}
With the same type of calculation, one can prove the following theorem:\\

\begin{theorem}\label{lambdastar} Let a channel have the property that 
\be\label{star}
\Lambda(\rho^*)=\Lambda(\rho)^*
\ee
where $*$ means complex conjugation. Then the complement of the channel has the following property
\be\label{starc}
\Lambda^c(\rho^*)=S^T\Lambda^c(\rho)^*S^*
\ee
where $S$ is a matrix which effects the following
\be\label{condstar}
K_\a^*=\sum_\beta S_{\a\beta}K_{\beta}.
\ee
\end{theorem}
\begin{proof}
Let the Kraus operators of the channel be as before. Property (\ref{star}) means that
\be
\sum_\a K_\a \rho^* K_\a^\dagger=\sum_\a K_\a^* \rho^* K_\a^T,
\ee
which is true only if the two sets of Kraus operators are related by a unitary transformation as in (\ref{condstar}). The correspondence (\ref{krauscomp})  then imposes the following condition
\be
(R_i)_{\a j}^*=\sum_\beta S_{\a\beta}^* (R_i)_{\beta j} 
\ee
which after inserting into the definition of the complementary channel and straightforward calculations proves (\ref{starc}).
\end{proof}
We now discuss the properties of the covariant Pauli channel and its complement. \\

\section{The Covariant Pauli channel and its symmetries}\label{Covpauli}
Consider first the general Pauli channel, defined as 
\begin{equation} \label{covpauli}
	\begin{split}
		\Lambda(\rho)=p_0 \rho +p_1 X\rho X +p_2 Y\rho Y+p_3 Z\rho Z,
	\end{split}     
\end{equation}
where $X,Y$ and $Z$ are the Pauli matrices and $p_0+p_1+p_2+p_3=1$. This channel is covariant under the discrete Pauli group, that is $\Lambda(g\rho g^\dagger)=g\Lambda(\rho)g^\dagger$, where $g\in \{\pm 1, \pm i\}\times  \{I,X,Y,Z\}$.
This covariance entails that the spectrum of the output state of the channel and its complement are invariant under inversions with respect to all the three axes, that is $Spect(\Lambda(\rho(x,y,z))=Spect(\Lambda(\rho(-x,-y,z))$ and similary for the other two axes.  
Moreover, the channel has also the symmetry (\ref{star}), which is due to the property $\sigma_i^*=\pm \sigma_i$. The combination of these two symmetries entails that the spectrum of the output state of the channel and its complement are invariant under the transformation ${\bf r}\lo -{\bf r}$. When $p_1=p_2$, which is the case under study in this work,  the channel (\ref{covpauli}), has a larger covariance under a continuous one-parameter group,
\begin{equation}
	\begin{split}
		\Lambda (U_z(\theta) \rho U_z(\theta) ^{\dagger})= U_z(\theta) \Lambda(\rho) U_z(\theta) ^{\dagger} \ \ \ \ \ \ \forall  \ U_z(\theta)   \in SO(2),
	\end{split}     
\end{equation}
where $U_z (\theta)=e^{\frac{i\theta Z}{2}}$. This extra covariance, then implies a further symmetry in the spectrum of the channel and its complement, namely that for any input state with Bloch vector ${\bf r}=(x,y,z)$, the eigenvalues of the output state of $\Lambda$ and its complementary depend only on $\sqrt{x^2+y^2}$ and $z$.\\

Furthermore if we temporarily denote this channel by $\Lambda_{p_0,p_3}$, it is readily seen that it has the nice property that   

\begin{equation}\label{change}
	\begin{split}
		\Lambda_{p_3,p_0}(\rho)= Z \ \Lambda_{p_0,p_3}(\rho) \ Z.
	\end{split}     
\end{equation}
 These symmetries will be important in the optimization which is needed for the calculation of capacities. Furthermore, the symmetry (\ref{change}) implies that all the capacities of the channel are symmetric with respect to the exchange of $p_0$ and $p_3$. 
For future reference, here we state the explicit form of the output state of the channel for an input state $\rho=\frac{1}{2}\left(\begin{array}{cc}1+z &x-iy\\ x+iy&1-z\end{array}\right)$:

\begin{equation}\label{channeloutput}
			\Lambda(\rho)=\frac{1}{2} 
		\begin{bmatrix} 
			1+(p_0+p_3-2p_1)z & (p_0-p_3)(x-iy) \\ (p_0-p_3)(x+iy)  & 1-(p_0+p_3-2p_1)z 
		\end{bmatrix}.
	 \end{equation}
The eigenvalues of this output state will be needed in the calculation of capacities, and are given by 
\begin{equation} \label{eigen}
	{\rm Spec}(\Lambda(\rho))=\frac{1}{2} (1\pm \sqrt{(p_0-p_3)^2 (x^2+y^2)+(2p_0+2p_3-1)^2 z^2)}.
\end{equation}
It is in order to find the complement of this channel and determine its symmetries. Using (\ref{krauscomp}), we find the Kraus operators of the complement of the covariant Pauli channel  
\begin{equation}
	R_1=\begin{bmatrix}
		\sqrt{p_0} & 0 \\ 0 & \sqrt{p_1} \\ 0& -i\sqrt{p_1} \\ \sqrt{p_3} & 0
	\end{bmatrix},
\end{equation}
\begin{equation}
	R_2=\begin{bmatrix}
		0&\sqrt{p_0}  \\  \sqrt{p_1}&0 \\ i\sqrt{p_1}&0 \\ 0& -\sqrt{p_3} 
	\end{bmatrix},
\end{equation}
which maps the same input state onto 
\begin{equation} \label{compch}
	\Lambda^c(\rho)=\begin{bmatrix}
		p_0 & \sqrt{p_0 p_1} x & \sqrt{p_0 p_1} y & \sqrt{p_0 p_3} z\\
		\sqrt{p_0 p_1} x & p_1 & -ip_1 z & i\sqrt{p_1 p_3} y \\
		\sqrt{p_0 p_1} y & ip_1 z & p_1 & -i\sqrt{p_1 p_3} x \\
		\sqrt{p_0 p_3} z & -i\sqrt{p_1 p_3} y & i\sqrt{p_1 p_3} x & p_3
	\end{bmatrix}.
\end{equation}
Relation (\ref{covofcomp}) now indicates that this channel has the covariance 
\begin{equation} \label{compcov}
	 \Lambda^c (\rho_{x',y',z})  =  \Omega^\dagger \Lambda^c (\rho_{x,y,z})\Omega,
\end{equation}
where $(x',y')=(x\cos\theta+y\sin\theta,-x\sin\theta+y\cos\theta)$ and 

\be
\Omega=\begin{bmatrix}
1 & 0 & 0 & 0 \\ 0 & \cos\theta & -\sin\theta & 0 \\ 0 & \sin\theta & \cos\theta & 0 \\ 0 & 0 & 0 & 1
\end{bmatrix}
\ee

\section{Capacities of the covariant Pauli channel} \label{capacitiysection}
For a quantum channel, one can define at least four different capacities \cite{gyongyosi2018survey}. These are the ultimate rates at which classical or quantum information can be transferred from a sender to a receiver per use of the channel by using different kinds of resources. There is a long route for converting these operational definitions to concrete and closed formulas for the capacities. Here we do not start from the operational definition for which the reader can refer to many good reviews \cite{gyongyosi2018survey, wilde2013quantum, lloyd1997capacity}, rather we start from the closed formulae which have been obtained for the calculation of capacity in each case \cite{schumacher1997sending, bennett2002entanglement, devetak2005capacity, devetak2005private}. Even after having these closed formulas, it is in general very difficult to find explicit values for the capacities in terms of the parameters of the channel. Besides super-additivity, the important property whose presence (or absence),  simplifies(or not) the calculation of some of these capacities  is the concavity of the relevant quantity which is to be maximized. We will see this in the following subsections, where we discuss the four capacities of the covariant Pauli channel.  

\subsection{Classical capacity}
\begin{figure}
     \centering
     \includegraphics[width=18.0 cm,height=14cm,angle=0]{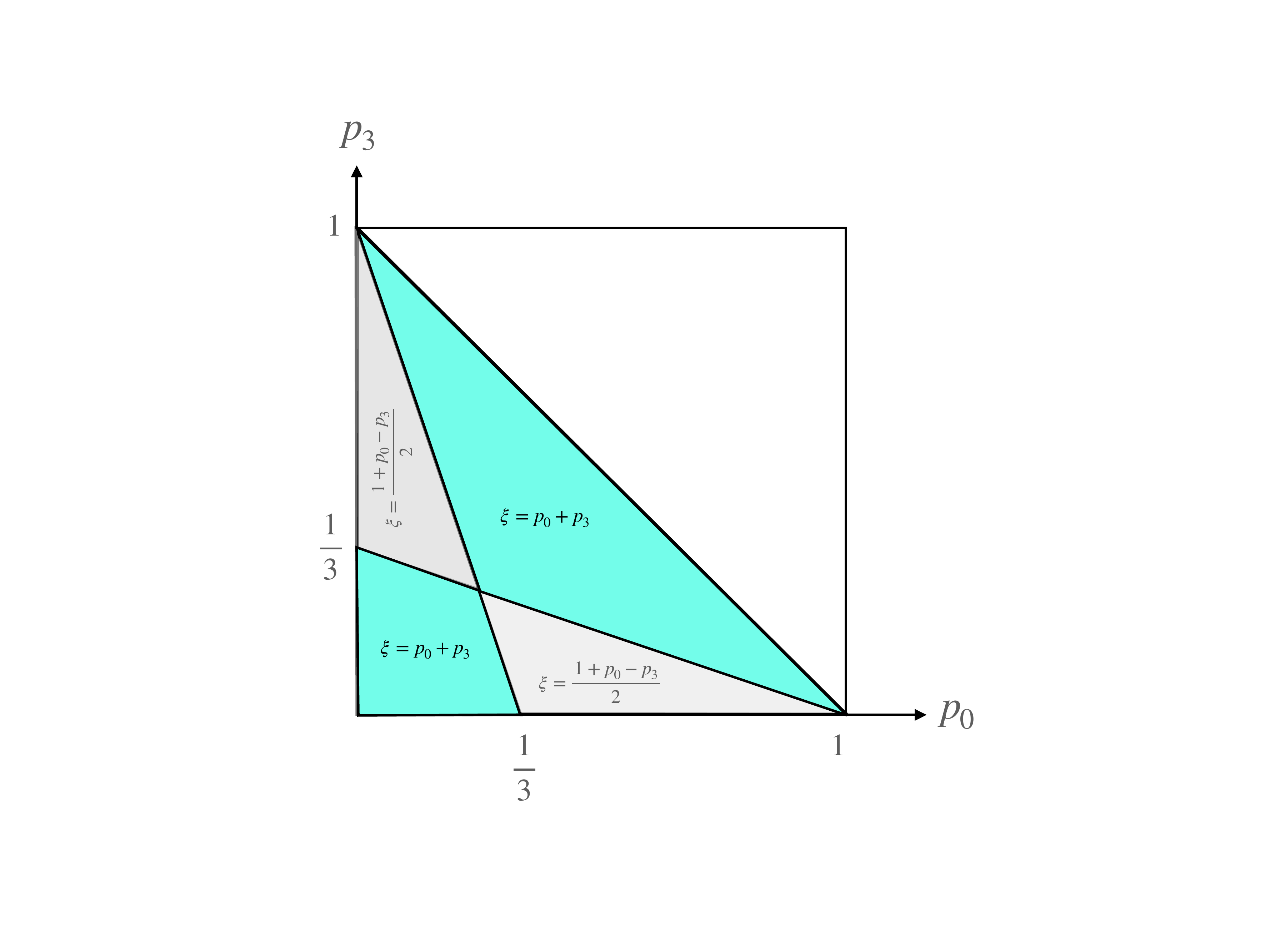}\vspace{-2cm}
     \caption{(Color Online) Two different regions for the Classical Capacity (\ref{split}, \ref{classcap}). In each region, the classical capacity is given $C_{cl}=1-h(\xi)$ where $h(\xi)$ is the binary entropy function.  }
     \label{ClassicalC}
 \end{figure}

 This is the ultimate rate that  classical messages, when encoded into quantum states, can be transmitted reliably over a channel.  It is given by  \cite{schumacher1997sending}: 
\begin{equation} \label{classicalcap}
	\begin{split}
		C_{cl}(\Lambda)= \lim_{n\longrightarrow \infty} \frac{1}{n} \chi^*(\Lambda^{\otimes n}),
	\end{split}   
\end{equation}
 where $\chi^*(\Lambda)= \max_{p_i , \rho_i} \ \chi\{p_i,\Lambda(\rho_i)\}$ \cite{wilde2013quantum} and $\chi\{p_i,\rho_i\}$ is the Holevo quantity of the output ensemble of states $\{p_i,\rho_i\}$ which is  defined as
\be
\chi(\{p_i,(\rho_i)\}):=S(\sum_i p_i\rho_i)-\sum_i p_iS(\rho_i).
\ee
Here $S(\rho)=-Tr(\rho\log \rho)$ is the von-Neumann entropy \cite{nielsen2002quantum}. 
In general $\chi^*$ is superadditive, meaning that $n\chi^*(\Lambda)\leq \chi^*(\Lambda^{\otimes n}) $, which makes the regularization in equation (\ref{classicalcap}) necessary for the calculation of the capacity \cite{hastings2009superadditivity}. However for unital qubit channels, which is the case at hand, it has been shown \cite{R17} that $\chi^*(\Lambda)$ is additive, and hence $C_{cl}(\Lambda)=\chi^*(\Lambda)$. It has been shown in \cite{wilde2013quantum} that one can only maximize $\chi\{p_i,\Lambda(\rho_i)\}$   over ensembles of pure input states.
 To find this ensemble, we proceed as follows. Let $|\phi\ra$ be a state which minimizes the output entropy, i.e. $S(\Lambda(\phi))\leq S(\Lambda(\rho))\ \ \ \forall \ \rho$.  Then in view of the explicit form of $\Lambda(\rho)$  in (\ref{channeloutput}) (or the invariance of its spectrum under ${\bf r}\lo -{\bf r}$), it is clear that $|\phi^\perp\rangle$ is also a minimum output entropy with exactly the same value.   Therefore we take the ensemble to be an equal mixture of $|\phi\ra$ and $|\phi^\perp\ra$. For this ensemble the first term of the expression of $\chi\{p_i,\Lambda(\rho_i)\}$ is maximized, since $\frac{1}{2}|\phi\ra\la\phi|+\frac{1}{2}|\phi^\perp\ra\la\phi^\perp|=\frac{I}{2}$ and $\Lambda$ being a unital channel maps this state to $\frac{I}{2}$. Therefore the problem reduces to finding the single state $|\phi\ra$ which minimizes the output entropy. In view of the $SO(2)$ covariance of the channel, we take the Bloch vector of this state to lie in the $xz$ plane, i.e. ${\bf r}=(\sin\theta,0,\cos\theta)$, then from (\ref{eigen}) we find the eigenvalues of the output state
\begin{equation}
	\lambda_{1,2}=\frac{1}{2} (1\pm\sqrt{ (p_0-p_3)^2 sin^2 \theta+(2p_0+2p_3-1)^2 cos^2\theta)}.
\end{equation}
Minimization of the output entropy is equivalent to maximizing the expression under the square root or the  function
\be
f(\theta)=(2p_0+2p_3-1)^2+\sin^2\theta \big[(p_0-p_3)^2-(2p_0+2p_3-1)^2\big].
\ee
This gives the optimum value of $\theta$ as
\be
\theta^{opt}=\begin{cases}
	\frac{\pi}{2}, & \text{if }  (p_0-p_3)^2 \ge (2p_0+2p_3-1)^2  \\
	0 , \pi, & \text{otherwise}.
\end{cases}
\ee
which leads to the expression of the capacity as 
\be \label{classcap}
C_{cl}(\Lambda)=1-h(\xi),
\ee
where $h(\xi)=-\xi \log\xi - (1-\xi)\log (1-\xi)$ is the binary entropy and 

\be \label{classcap1}
\xi=\begin{cases}
	\frac{1+p_0-p_3}{2} & \text{if }  (p_0-p_3)^2\geq (2p_0+2p_3-1)^2  \\
	p_0+p_3 & \text{otherwise}.
\end{cases}
\ee
The two regions are shown in figure (\ref{ClassicalC}). 

\subsection{Entanglement-assisted capacity}
This is the ultimate rate for transmitting classical information over a quantum channel when the sender and receiver share an unlimited source of entanglement. Fortunately, it has been proved that finding this type of capacity does not need a regularization and only needs a finite and convex optimization \cite{bennett2002entanglement}:
\begin{equation}
	\begin{split}
		C_E(\Lambda)=\max_\rho I(\rho , \Lambda),
	\end{split}     
\end{equation}
where
\begin{equation}
	I(\rho , \Lambda)= S(\rho) + S (\Lambda(\rho))-S(\Lambda^c (\rho)),
\end{equation}
 is the quantum mutual information.
The eigevalues of $\rho$, $\Lambda(\rho)$ and $\Lambda^c(\rho)$ are all invariant under the transformation 
Covariance and symmetries of the channel, leads to symmetries of the spectrum of output states under the transformation ${\bf r}\lo {\bf -r}$. In conjunction with the concavity of the mutual information $
 \lambda_1I(\rho_1,\Lambda)+\lambda_2I(\rho_2,\Lambda)\leq I(\lambda_1\rho_1+\lambda_2\rho_2,\Lambda)
 $, this implies that the optimal state is nothing but the completely mixed state. In fact from the symmetry it is evident that if a state $\rho(\bf r)$ maximizes the mutual information, then the state $\rho(-{\bf r})$ also does, and  from concavity one finds that the maximally mixed state  $\frac{1}{2}(\rho({\bf r})+\rho({\bf -r}))=\frac{I}{2}$, is the global maximum of the mutual information. This gives
\begin{equation}\label{Cent}
			C_E(\Lambda)=I(\frac{I}{2}, \Lambda)=2+p_0 log \ p_0 + (1-p_0-p_3) log\ \frac{1-p_0-p_3}{2} +  p_3 log \ p_3.
\end{equation}

\subsection{Quantum capacity}
This is the ultimate rate for transmitting quantum information and preserving the entanglement between the channel's input and a reference quantum state over a quantum channel.  This quantity is described in terms of coherent information \cite{devetak2005private}:
\begin{equation}
	\begin{split}
		C_q(\Lambda) =
		\lim_{n \to \infty}  \; \frac{1}{n} J(
		\Lambda^{\otimes n}),
	\end{split}     
\end{equation}
where $J(\Lambda)=\max_{\rho} J(\rho,\Lambda)$ and $J(\rho,\Lambda):=S(\Lambda(\rho))-S(\Lambda^c(\rho))$. It is known that $J$  is superadditive, i.e. $J(\Lambda_1 \otimes \Lambda_2)\ge J(\Lambda_1)+J(\Lambda_2)$, rendering an exact calculation of the quantum capacity extremely difficult and at the same time providing a lower bound in the form  $C_q^{(1)}(\Lambda)\leq C_q(\Lambda)$ where $C_q^{(1)}:=J(\Lambda)$ is the single shot capacity.  However if the channel is  degradable, then the additivity property is restored $C_q(\Lambda)=C_q^{(1)}(\Lambda)$ \cite{devetak2005capacity}, and the calculation of the quantum capacity becomes a convex optimization problem. \\ 

 For a channel like the covariant Pauli channel in (\ref{covpauli}), where we know that is not degradable \cite{cubitt2008structure}, we have to be content with partial results. In what follows we  proceed to find partial and yet important  information on the quantum capacity. First we determine the regions where the channel has zero capacity and then in the complement of this region, we find lower and upper bounds for the quantum capacity. 
 
\subsubsection{The zero capacity region}
\begin{figure}
     \centering
     \includegraphics[width=18.0 cm,height=14cm,angle=0]{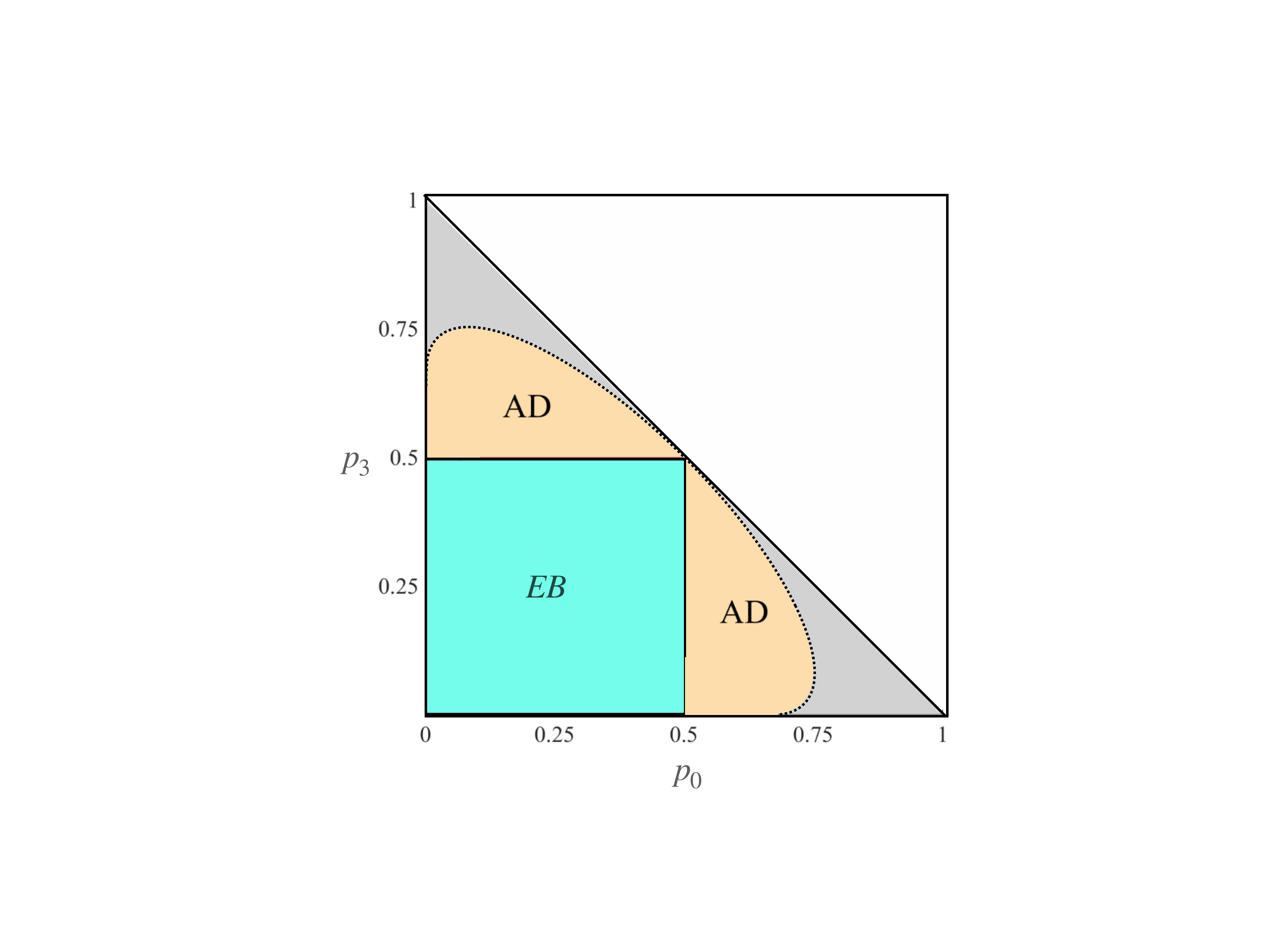}\vspace{-2.1cm}
     \caption{ (Color Online) Zero capacity regions: the entanglement breaking (EB) and anti-degradable (AD) regions are shown in the parameters space. The curved dotted line indicates the solution of inequality (\ref{myeq}).}
     \label{zerocapacityregion}
 \end{figure}

It is known that Entanglement breaking channels cannot  preserve quantum correlations and thus their quantum capacity is equal to zero \cite{horodecki2003entanglement}. We remind the reader that an Entanglement Breaking channel (EB)  is one which when acting on a state, breaks its entanglement with any other state. It is known that a channel $\Lambda$ is entanglement breaking if its Choi matrix defined as  ${\cal J}_{\Lambda}=2(\Lambda\otimes I)|\phi^+\ra\la \phi^+|$ is a separable state for $|\phi^+\ra$ a maximally entangled state $|\phi^+\rangle=\frac{1}{\sqrt{2}} (|00\rangle +|11\rangle)$. For checking this property, we use the Peres criteria \cite{peres1996separability} to determine the region where the following state is separable: 
\begin{align}
	{\cal J}_{\Lambda}=2\Lambda \otimes I (|\phi^+\rangle\langle\phi^+|)= 
	\begin{bmatrix}
		p_0 + p_3 & 0 & 0 & p_0-p_3 \\
		0 & 2p_1 & 0 & 0 \\
		0 & 0 & 2p_1 & 0 \\ 
		p_0-p_3 & 0 & 0 & p_0+p_3
	\end{bmatrix}.
\end{align}
The partial transpose of this matrix is  
\begin{align}
	(I\otimes T){\cal J}_{\Lambda}	= 
	\begin{bmatrix}
		p_0 + p_3 & 0 & 0 & 0 \\
		0 & 2p_1 & p_0-p_3 & 0 \\
		0 & p_0-p_3& 2p_1 & 0 \\ 
		0 & 0 & 0 & p_0+p_3
	\end{bmatrix}.
\end{align}
whose eigenvalues (in view of the normalization $p_0+2p_1+p_3=1$) are given by
\be
\{p_0+p_3, p_0+p_3,1-2p_3,1-2p_0\}.
\ee
This shows that in the region $$(p_0\leq \frac{1}{2}) \wedge (p_3\leq \frac{1}{2}),$$ the capacity of the channel is zero. This region is shown in figure (\ref{zerocapacityregion}).  It turns out however that the zero-capacity region is slightly larger than this. To this end we use another theorem \cite{smith2012detecting} according to which any channel which is anti-degradable, has zero capacity\cite{smith2012detecting}. To check the anti-degradability of $\Lambda$, we use a criterion first proposed in \cite{paddock2017characterization}. According to this criterion, a channel is anti-degradable, if 
\begin{equation} \label{anticon}
tr({\cal J}_{\Lambda}) ^2- 4\sqrt{det({\cal J}_{\Lambda})}\leq 	tr(\Lambda(I)^2),
\end{equation}
where ${\cal J}_{\Lambda}= d(\Lambda \otimes I) |\phi\rangle\langle\phi|$ is the choi matrix. In our case:
\begin{equation}
	{\cal J}_{\Lambda}= \begin{bmatrix}
		p_0 + p_3 & 0 & 0 & p_0-p_3 \\
		0 & 2p_1 & 0 & 0 \\
		0 & 0 & 2p_1 & 0 \\ 
		p_0-p_3 & 0 & 0 & p_0+p_3
	\end{bmatrix},
\end{equation}
 whose   trace and determinant are respectively given by $tr({\cal J}_{\Lambda}^2)= 4p_0 ^2 + 4 p_3 ^2 + 8 p_1 ^2$, and  $det({\cal J}_{\Lambda})= 16 \ p_0 p_3 p_1 ^2$. Using (\ref{anticon}), the anti-degradability condition becomes:
\begin{equation}\label{myeq}
 2(p_0 ^2+p_3 ^2)+(1-p_0-p_3)^2-4(1-p_0-p_3)\sqrt{p_0 p_3}\leq 1.
\end{equation}
The region where this inequality holds is shown in figure (\ref{zerocapacityregion}) which includes the Entanglement-Breaking region.\\

\subsubsection{Flag extension and upper bounding the quantum capacity}
Outside the zero-capacity region, the quantum capacity can only be upper bounded. 
 A novel technique for upper bounding the quantum capacity is flag extension which has been successfully used for investigating  various channel's quantum capacities \cite{fanizza2020quantum}\cite{kianvash2020bounding}\cite{fanizza2021estimating}. For any channel written as the convex combination of other channels, 
\begin{equation}
	\Lambda(\rho)=\sum_i p_i \Lambda_i (\rho),
\end{equation}
we can construct a new channel with a higher dimensional output Hilbert space:
\begin{equation}
	\Lambda_f(\rho)=\sum_i p_i \Lambda_i (\rho) \otimes \sigma_i,
\end{equation}
where $\sigma_i \in L^+(H_F)$ are the flag states,  $H_F$ is called the flag space, and $\Lambda_f$ is called the flagged channel. Since $tr_F \Lambda_f = \Lambda$, one has: 
\begin{equation}
	C_q(\Lambda) \leq   C_q(\Lambda_f).
\end{equation}
Therefore by constructing a flag extension for the channel which is degradable, one can find the quantum capacity of the flagged channel and upper bound the capacity of the original channel. In this regard, the following theorem is notable \cite{smith2008additive}: \\

\begin{theorem} \cite{smith2008additive}\label{thm1}  Suppose we have the channel 
\begin{equation}
	\Lambda(\rho)=\sum_i p_i \Lambda_i (\rho),
\end{equation}
where all the $\Lambda_i$'s are degradable channels. Then the flag extension 
\begin{equation}
	\Lambda_f(\rho)=\sum_i p_i \Lambda_i (\rho) \otimes |i\rangle\langle i|,
\end{equation}
is a degradable channel and: 
\begin{equation} \label{upperbound}
	C_q(\Lambda)\le C_q(\Lambda_f)= \sum p_i C_q(\Lambda_i).
\end{equation}
\end{theorem}
\noindent To use this theorem we need an appropriate convex decomposition of the covariant Pauli channel. One possible decomposition is: 
\begin{equation}\label{decomp1}
	\Lambda= (p_0+p_3) \Lambda_0 + 2p_1 \Lambda_1,
\end{equation}
where
\begin{equation}
	\Lambda_0 (\rho)=\frac{1}{p_0+p_3} (p_0 \rho+ p_3 Z\rho Z) ,
\end{equation}
and
\begin{equation}
	\Lambda_1 (\rho)=\frac{1}{2} (X \rho X+ Y\rho Y) .
\end{equation}
First, we  prove that these channels are degradable and then we use theorem (\ref{thm1}) to obtain an upper bound for the quantum capacity.  
We can analyse the degradability of $\Lambda_0$ indirectly by checking the anti-degradability of it's complementary channel. The output state of the channel and its complementary  and the  Choi matrix of the complementary channel are respectively:
\begin{equation} \label{lambda0}
	\Lambda_0(\rho)= \frac{1}{2} \begin{bmatrix}
		1+z & \frac{p_0-p_3}{p_0+p_3}(x-iy)  \\ \frac{p_0-p_3}{p_0+p_3}(x+iy) & 1-z \end{bmatrix},
\end{equation}

\begin{equation} \label{lambdaa}
	\Lambda_0 ^c(\rho)= \begin{bmatrix}
		\frac{p_0}{p_0+p_3} & \frac{\sqrt {p_0 p_3}}{p_0+p_3} z  \\ \frac{\sqrt {p_0 p_3}}{p_0+p_3} z  &\frac{p_3}{p_0+p_3} \end{bmatrix},
\end{equation}

\begin{equation}
	{\cal J}_{\Lambda_0 ^c}= \begin{bmatrix}
		\alpha^2& \alpha \beta & 0 & 0 \\ \alpha \beta & \beta^2 & 0 & 0 \\ 0 & 0 & \alpha^2 & -\alpha \beta \\ 0 & 0 & -\alpha \beta & \beta^2
	\end{bmatrix},
\end{equation}
where $\alpha=\sqrt{\frac{p_0}{p_0+p_3}}$ and $\beta=\sqrt{\frac{p_3}{p_0+p_3}}$. One can easily see that $tr(\Lambda_0 ^c(I)^2)=4\frac{p_0^2+p_3^2}{(p_0+p_3)^2}$, $tr({\cal J}_{\Lambda_0 ^c}^2)=2$ and $det({\cal J}_{\Lambda_0 ^c})=0$. The condition (\ref{anticon}), reduces to the  valid inequality $(p_0-p_3)^2 \ge 0$. This shows that $\Lambda_0 ^c$ is anti-degradable which implies that  $\Lambda_0$ is degradable. Also, since the channel $\Lambda_1$ is unitarily equivalent to $\Lambda_0(p_0=p_3)$, (i.e. $\Lambda_1=X \Lambda_0(p_0=p_3) X$, this channel is also degradable. Therefore the flag extension 
\be \Lambda_f= (p_0+p_3) \Lambda_0 \otimes|0\rangle\langle0|+2p_1 \Lambda_1 \otimes|1\rangle\langle1|,\ee
 is a degradable extension and eq. (\ref{upperbound}) applies to our channel. The single-shot quantum capacity of $\Lambda_0$ is found from 
\begin{equation}
	C_q(\Lambda_0)= \max_{\rho} \big[ S(\Lambda_0 (\rho))-S(\Lambda_0 ^c (\rho))\big].
\end{equation}
The explicit form of $\Lambda_0(\rho)$ and $\Lambda_0^c(\rho)$ in (\ref{lambda0}) and (\ref{lambdaa}) now show that the output states of these channels are invariant under the transformation $(x,y,z)\longrightarrow (-x,-y,-z) $ which implies that the output state is  $\rho_{opt}=\frac{I}{2}$, leading to
\begin{equation}
	C_q(\Lambda_0)= S(\Lambda_0(I/2))-S(\Lambda_0^c (I/2))=1-h(\frac{p_0}{p_0+p_3}),
\end{equation}
where $h(x)=-x log(x)-(1-x) log(1-x)$ is the binary entropy. 
Also, this calculation directly shows that $C_q(\Lambda_1)=0$. So, using (\ref{upperbound}), an analytic upper bound is  derived for the covariant Pauli channel:
\begin{equation} \label{q1}
	C_q(\Lambda)\le (p_0+p_3)C_q(\Lambda_0)=(p_0+p_3)(1-h(\frac{p_0}{p_0+p_3}))=:A.
\end{equation}
One can write this in a more informative way by redefining $\epsilon:=\frac{p_0-p_3}{p_0+p_3}$ which indicates the amount of asymmetry of the channel. This leads to the following expression
\be \label{uppercov}
C_q(\Lambda)\le (p_0+p_3) C_q(\Lambda_0):=(p_0+p_3)(1-h(\frac{1+\epsilon}{2})).
\ee

 \begin{figure}
     \centering
     \includegraphics[width=15.4 cm,height=11.3cm,angle=0]{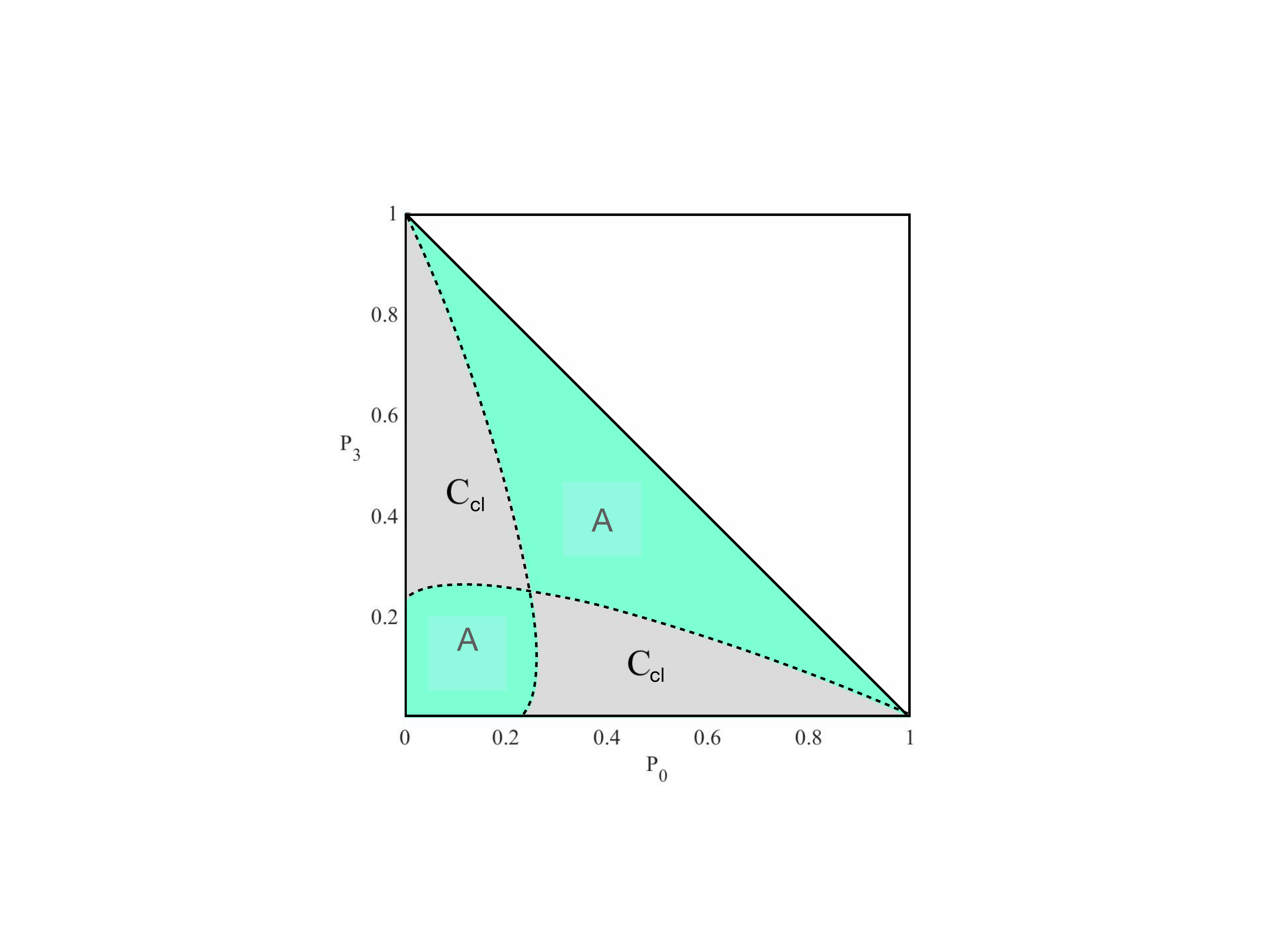}\vspace{-2.0cm}
     \caption{(Color Online) Two different regions for the best upper bound. In one region the best upper bound is given by $A$ as in equation (\ref{q1}) or (\ref{uppercov}) and in the other region, the classical capacity $C_{cl}(\Lambda)=1-h(\xi)$, with $\xi=\frac{1+p_0-p_3}{2}$ is the best upper bound. The curved dotted boundaries depicts the solution of equation (\ref{cq1}). }
     \label{Compareqc}
 \end{figure}
  Incidentally we note that for $\epsilon=0$, the upper bound is 0, which shows that the symmetric Pauli channel $\Lambda(\rho)=p_0(\rho+Z\rho Z)+p_1(X\rho X+Y\rho Y)$ has zero capacity. This is in accord with our previous result on zero-capacity region shown in figure (\ref{zerocapacityregion}).\\
    \noindent Note however that (\ref{decomp1}) is not the only convex decomposition of the channel. The channel has indeed another decomposition, the other one being:
  \begin{equation}
\Lambda= (p_0+p_1) \Lambda_2 + (p_1+p_3) \Lambda_3,
\end{equation}
where: 
\begin{equation}
\Lambda_2 (\rho)=\frac{1}{p_0+p_1} (p_0 \rho+ p_1 X\rho X),
\end{equation}

\begin{equation}
\Lambda_3 (\rho)=\frac{1}{p_1+p_3} (p_1 Y\rho Y+ p_3 Z\rho Z).
\end{equation}
  The output of $\Lambda_2$ and its complementary channel are respectively:
  
  \begin{equation} \label{lambda2}
\Lambda_2 (\rho)= \frac{1}{2(p_0+p_1)}\begin{bmatrix}
1+(p_0-p_1)z & (p_0+p_1)x-i (p_0-p_1)y \\ (p_0+p_1)x+i (p_0-p_1)y & 1-(p_0-p_1)z
\end{bmatrix}.
\end{equation}

\begin{equation} \label{lambda2c}
\Lambda_2 ^c (\rho)= \frac{1}{p_0+p_1}\begin{bmatrix}
p_0 & \sqrt{p_0 p_1} x \\ \sqrt{p_0 p_1} x & p_1
\end{bmatrix},
\end{equation}
By the same procedure followed  for $\Lambda_0$, the degradability condition of $\Lambda_2$ reduces to the condition $(p_0-p_1)^2 \ge 0$ which shows that $\Lambda_2$ and thus its equivalent channel $\Lambda_3$ ($\Lambda_3= Y \Lambda_2(p_0=p_1,p_1=p_3) Y$) are both degradable. According to the invariance of the spectrum of (\ref{lambda2}) and (\ref{lambda2c}) with respect to $(x,y,z)\longrightarrow (-x,-y,-z)$ the optimal state is the completely mixed state and: 
\begin{equation} 
C_q(\Lambda_2)= (1-h(\frac{p_1}{p_0+p_1})),
\end{equation}

\begin{equation}
C_q(\Lambda_3)= (1-h(\frac{p_1}{p_3+p_1})).
\end{equation}
This leads to the upper bound:
\begin{eqnarray} \label{q2}
C_q(\Lambda)&\le& (p_0+p_1)C_q(\Lambda_2)+(p_1+P_3)C_{q}(\Lambda_3)\cr &=& (p_0+p_1)(1-h(\frac{p_1}{p_0+p_1}))+ (p_1+p_3)(1-h(\frac{p_1}{p_3+p_1}))=:B.
\end{eqnarray}
Before comparing $A$ and $B$, we note that the classical capacity $C_{cl}$ is also an upper bound on the quantum capacity, so the  upper bound for the quantum capacity is the following:

\begin{equation} \label{Qbest}
C_q(\Lambda) \le  \min \{A,B, C_{cl}\},
\end{equation}
where $A, B$ and $C_{cl}$ are given respectively in (\ref{q1}), (\ref{q2}) and (\ref{classcap}).
Figure (\ref{Compareqc}) shows the region where one of these capacities is the upper bound. It is seen that $B$ is never lower than the other two capacities and hence the comparison is between $A$ and $C_{cl}$. The regions are almost similar but not identical to those of exact classical capacity. The curved boundary is the solution of the equation $C_{cl}=A$ or more explicitly the following equation:

\begin{equation}\label{cq1}
    (p_0+p_3)(1-h(\frac{p_0}{p_0+p_3}))=1+ (p_0+p_3)log \ (p_0+p_3) + (1-p_0-p_3)  log \ (1-p_0-p_3),
\end{equation}

 \begin{figure}
     \centering
     \includegraphics[width=18.4 cm,height=11.5cm,angle=0]{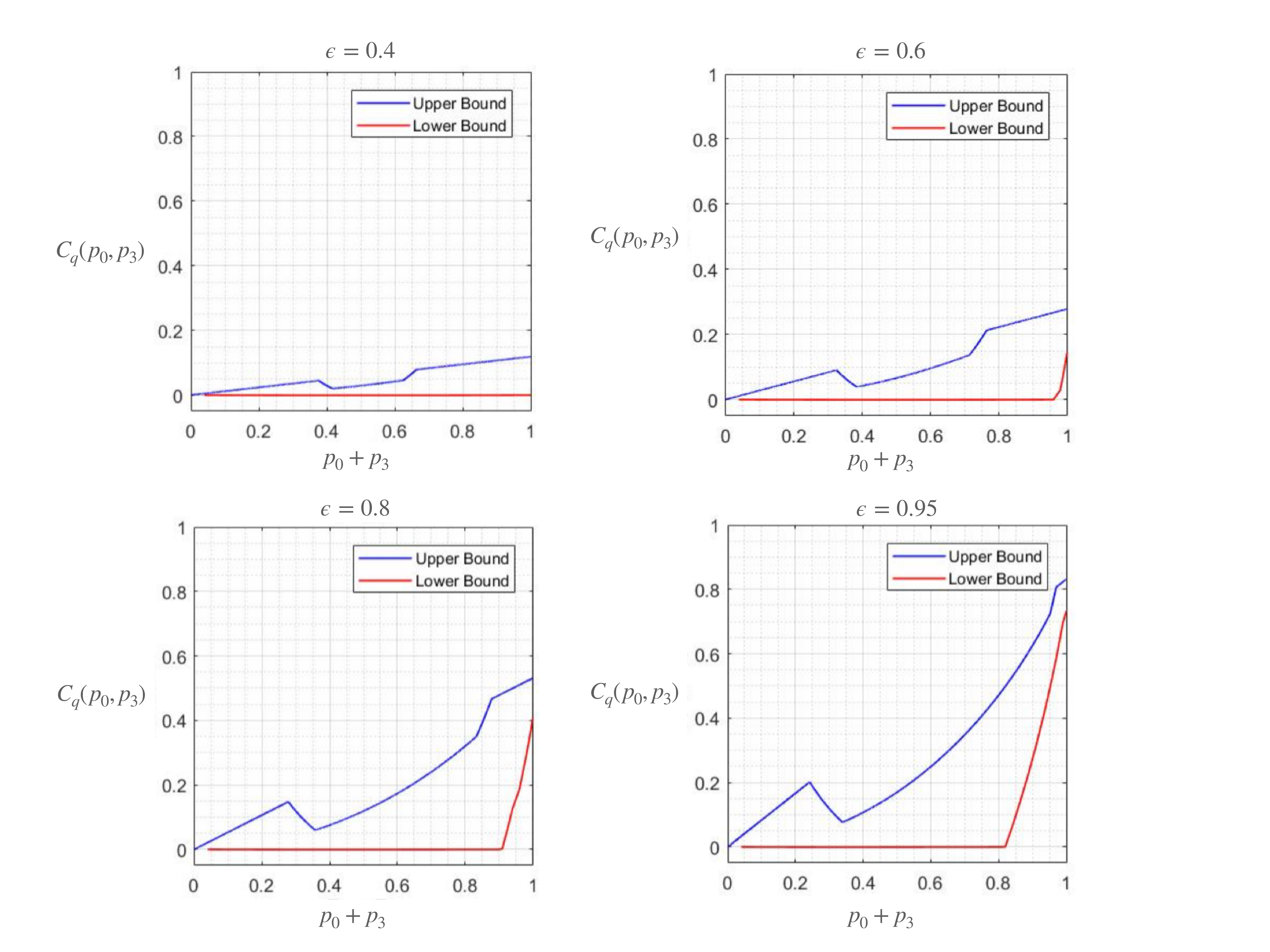}\vspace{-0.5cm}
     \caption{(Color Online) In all the figures, the upper curves show the best upper bound (eq.\ref{Qbest}) and the lower curves show the best lower bounds (eq.\ref{lower}) of quantum capacity vs. $p_0+p_3$ for fixed values of $\epsilon:=\frac{p_0-p_3}{p_0+p_3}$}
     \label{uplow}
 \end{figure}
  
\subsubsection{Lower bounding the Quantum capacity}
A natural lower bound for the quantum capacity is the single shot capacity 
\begin{equation}
	C_q^{(1)}(\Lambda)= \max_{\rho} J(\rho , \Lambda),
\end{equation}
where $J(\rho,\Lambda)=S(\Lambda(\rho))-S(\Lambda^c(\rho))$. Due to the superadditivity of the coherent information of the channel, we know that 
$C_q^{(1)}(\Lambda)\leq C_q(\Lambda)$. The problem however is that for $\Lambda$, which is non-degradable, the coherent information is not concave and the covariance of the channel does not entail a simple form for the optimal state, as it did for the degradable channels $\Lambda_0$ and $\Lambda_1$. Therefore we have to resort to numerical calculations. \\

\noindent The eigenvalues of the channel's output are given in (\ref{eigen}) which due to the covariance, depend only on $r=\sqrt{x^2+y^2}$. The same is also true for the eigenvalues of the complementary channel (\ref{compch}), due to the induced covariance (\ref{compcov}). Denoting the eigenvalues of $\Lambda^c(\rho)$ by $\mu_i$, we can search for the optimum state among all Bloch vectors which have the same value of $r=\sqrt{x^2+y^2}$. The expression that we shall maximize is 
 \begin{equation}
	J=-\sum_{i=1}^{2} \lambda_i log \ \lambda_i + \sum_{i=1}^{4} \mu_i log \ \mu_i,
\end{equation}
and the  maximization  needs to be done over $r$ and $z$: 
\begin{equation} \label{lower}
	C_q^{(1)}(\Lambda)=\max_{r,z} J(p_0,p_3,r, z).
\end{equation}
By doing this optimization numerically, we obtain a lower bound for the quantum capacity of the channel. Figure (\ref{uplow}) shows this the upper and lower bound for various values of $\epsilon$  as functions of $p_0+p_3$.

\subsection{Private capacity}
This is the ultimate rate for transmitting classical information encoded in quantum states without leaking information to the environment \cite{devetak2005private}: 
\begin{equation}
\begin{split}
     C_p= \lim_{n\longrightarrow \infty} \frac{1}{n} P(\Lambda^{\otimes n}),
\end{split}     
\end{equation}
with $P(\Lambda)=\max_{p_i,\rho_i} (\chi\{p_i,\Lambda(\rho_i)\}-\chi\{p_i,\Lambda^c(\rho_i)\})$. As proved in \cite{li2009private}, this quantity is super-additive and can not be obtained explicitly. In comparison with quantum capacity, $C_p(\Lambda)\ge C_q(\Lambda)$ but for degradable channels we know that $C_p(\Lambda)= C_q(\Lambda)$ \cite{devetak2005capacity}. So, the upper bound for the quantum capacity of the channel (\ref{uppercov})is also a valid upper bound for the private capacity. \\

\section{Discussion}
The theory of quantum communication is much richer than that of classical communication, due to the possibility of entangled input states, entangled measurements and shared entanglement between the sender and the receiver. This opens up completely new phenomena, like superadditivity, superactivation and causal activation which have no counterpart in the classical world \cite{calefi1, calefi2}. In the present paper, our aim has been very limited. We have tried to study the four 
conventional capacities of a specific class of quantum channels which are of great interest. This is the family of of Pauli channels, defined by 
\begin{equation}
\Lambda (\rho)=p_0\rho + p_1(X\rho X+Y\rho Y)+p_3 Z\rho Z,\ \ \ \ p_0+2p_1+p_3=1,
\end{equation}
\noindent which we call the Covariant Pauli channel, due to its covariance with respect to the group of rotations around the $z-$ axis, Eq.(5) and its $Z_2$ symmetry, Eq.(24).
These family of channels include the depolarizing channel as a subset for which  similar studies have been carried out in \cite{fanizza2020quantum}. This study therefore is an intermediate step toward the study of 
the full Pauli channel. \\

 We were interested in four conventional types of capacities of this channel which needed entropic optimizations. These are the classical capacity $C(\Lambda)$, entanglement assisted capacity $C_E(\Lambda)$, quantum capacity $Q(\Lambda) $ and private capacity $C_p(\Lambda)$. 
 As the complementary channel's output entropy appears in most of these optimizations, we proved a general theorem, theorem 1, on the connection between the covariance properties of a channel and its complement. This theorem may find applications in other similar studies, even in contexts other than calculation of capacities.  Using this theorem and invoking the properties of unitality, covariance and symmetry of the channel, we could  \\
 
 a-calculate exact expressions for $C_{cl}(\Lambda)$, Eq. (\ref{classcap}), \\
 
 b-calculate an exact expression of  $C_E(\Lambda)$, Eq. (\ref{Cent}),\\
 
 c-find lower and upper bound for the $C_q(\Lambda)$ and $C_p(\Lambda)$, Figs. (\ref{uplow}),\\
 
 d-determine large regions in the parameter space, where $C_q(\Lambda)$ is exactly zero, Fig. (\ref{Compareqc}).\\
 
 Usually it is extremely difficult or impossible to calculate capacities of quantum channels due to the super-additivity problem,  However in certain cases, the super-additivity obstacle is alleviated. This is the case for the classical capacity of unital channels  and the entanglement assisted capacity, which we invoked for (a) and (b).  
 The important quantum capacity $C_q(\Lambda)$ however lies beyond exact calculation due to the super-additivty issue. Therefore for the quantum capacity, we used flag-extension technique (\cite{kianvash2020bounding}) and invoked the covariance of the channel and its complement, combined with a brute force search to obtain upper and lower bounds, shown in figure (\ref{uplow}).     Finally we used a theorem in (\cite{smith2012detecting}) to find the regions where the channel is anti-degradable and hence its quantum capacity is exactly zero. 
 It will be interesting to extend these results to the general Pauli channel in higher dimensions. Also, one can slightly deform the Landau-Streater channel 
 $$
 \Lambda_{LS}(\rho)=\frac{1}{j(j+1)}(J_x\rho J_x+J_y\rho J_y+J_z\rho J_z)
 $$
 where $J_a, a =x,y,z$ are the spin $j$ rotation operators
 \cite{LandauStreater}. The quantum informational properties, including degradibility and two types of capacities of this channel, which has full $SU(2)$ covariance, has already been studied in (\cite{filippov2019quantum}). One can slightly deform this channel and retain its covariance with respect to a subgroup of $SU(2)$, and make similar studies as in the present work.  
 Furthermore, as figure (\ref{Compareqc}) shows, the covariant Pauli channel is degradable on the line $p_0+p_3=1$. This means that near this line, the channel may be approximately degradable,\cite{approximate}, implying  that there exists a degrading channel $\Phi$ which upon composition with the channel $\Lambda$ is $\epsilon$-close in the diamond norm to the complementary channel $\Lambda^c$, i.e. $\left\Vert \Lambda^c-\Phi\circ \Lambda\right\Vert_{\Diamond}\leq \epsilon$.  Therefore, it may be possible to obtain slightly better bounds for the quantum capacity near the line $p_0+p_3=1.$ We leave these investigations to a future publication.

\section{acknowledgments} We thank anonymous referees for their very valuable comments for the improvement of this work. We acknowledge partial financial support of Iran National Science Foundation under the grant INSF-98024071. We would like to thank the members of the Quantum Information Group in Sharif for many useful discussions. Abbas Poshtvan would like to thank Ali Sourchap for his contribution in obtaining numerical results, Amir Arqand and Farzad Kianvash for useful discussions in the early stages of this project and Hoda Masoud for her help in designing the figures.

\section{References}
\bibliography{apssamp}
\end{document}